\documentclass{ws-procs9x6}

\begin{document}

\title{Obtaining Parton Distribution Functions from Self-Organizing Maps\footnote{\uppercase{W}ork 
supported by grants 0426971 of the \uppercase{US} 
\uppercase{N}ational \uppercase{S}cience \uppercase{F}oundation and 
and \uppercase{DE-FG}02-01\uppercase{ER}41200 of the \uppercase{US} \uppercase{D}epartment of \uppercase{E}nergy.}}

\author{H. HONKANEN\footnote{hh9e@virginia.edu}  and S. LIUTI\footnote{sl4y@virginia.edu}}

\address{Physics Department \\ 382 McCormick Rd., University of Virginia 
\\ Charlottesville, Virginia 22904, USA}

\author{Y.C. LOITIERE\footnote{ycl2r@virginia.edu}, D. BROGAN\footnote{dbrogan@virginia.edu} and P. REYNOLDS\footnote{reynolds@virginia.edu}}

\address{Department of Computer Science \\
School of Engineering, University of Virginia \\
151 Engineer's Way, P.O. Box 400740 \\
Charlottesville, Virginia 22904 USA}  

\maketitle

\abstracts{We present an alternative algorithm to global fitting procedures  
to construct Parton Distribution Functions parametrizations. 
The proposed algorithm uses Self-Organizing Maps which at variance with 
the standard Neural Networks, are based on competitive-learning. 
Self-Organizing Maps generate a non-uniform projection from a high dimensional data space
onto a low dimensional one (usually 1 or 2 dimensions) by clustering 
similar PDF representations together. 
The SOMs are trained on 
progressively narrower selections of data samples.
The selection criterion is that of convergence towards a neighborhood of
the experimental data. 
All available data sets on deep 
inelastic scattering 
in the kinematical region of $0.001 \leq x \leq 0.75$, 
and $1 \leq Q^2 \leq 100$
GeV$^2$, with a cut on the final state invariant mass, $W^2 \geq 10$ GeV$^2$
were implemented.
The proposed fitting procedure, at variance
with standard neural network approaches,  
allows for an increased control of the systematic  
bias by enabling the user to directly control the data 
selection procedure at various stages of the process.
}

\section{Introduction}
Parton Distribution Functions (PDFs) are defined as the probabilities
to find a parton -- a quark, antiquark or a gluon -- of type $a$ in the 
proton with a given value of the process' scale defined by $Q^2$, the  
four-momentum transfer squared, and Bjorken's variable, $x_{Bj}=Q^2/2M\nu$, 
$\nu$ being the energy transfer and $M$ the proton mass. $x_{Bj}$ represents
the light-cone momentum fraction of the proton carried by the parton.
Although PDFs were studied both theoretically and experimentally for the past few decades, 
their determination is still hampered by a number of unsolved questions 
mainly concerning their Perturbative QCD (PQCD) evolution and, related to this, 
the treatment of heavy flavor quarks.
Furthermore, this situation -- in particular the large indetermination of the gluon distribution --
will have practical critical consequences on the predictivity of results 
at the LHC. PDFs were, in fact, 
recently defined as ``a necessary evil'' \cite{Pumplin:2005yf}.   
Our work was indeed motivated by similar concerns as the ones expressed
in \cite{Pumplin:2005yf}. 

To date, a few approaches have been developed that deal with the question of a 
fully quantitative determination of PDFs in a wide range of $x_{Bj}$ and $Q^2$. On one
side we have {\em Global Fitting} (GF) procedures, pursued, developed and 
refined since the beginning of QCD. 
\footnote{All results by the active groups in recent years are listed in 
\protect\cite{durham}, 
and are also reported regularly at this conference.}  
More recently, a number of alternative approaches to GF were pursued, 
the main ones being the {\em Neural Network} (NN) approach \cite{Piccione}, 
and the Bayesian methods \cite{cowan}.  
In both Refs.\cite{Piccione,cowan}, the authors are concerned with the definition 
and evaluation of the PDFs uncertainties from GF. 
In particular, the $\chi^2$ obtained from the GF procedure is 
most likely to underestimate both the theoretical and experimental 
errors from the various data sets as proven by the existence of often large 
discrepancies in the results obtained by 
different groups \cite{Pumplin:2005yf}. 
In Ref.\cite{Piccione}, in particular, the main source
of indetermination is attributed to the {\em theoretical bias} 
introduced by the choice of parametrization form
of PDFs at the initial scale, $Q_o^2$, of PQCD evolution. 
However, implicit in NN algorithms is a hardly controllable {\it systematic bias}. 
The approach we propose here is based on a specific class of neural network algorithms, 
the Self-Organizing Maps (SOMs) (for a review  see \cite{Kohonen}). 
SOMs allow for a better control of the 
systematic bias by allowing to replace the fully automated procedure of standard NNs with 
an interactive fitting procedure, at the 
expense of re-introducing some theoretical bias in the fit.
Our fitting procedure is based on an iterative 
process in which the ``user'' interactively delineates the boundary 
between acceptable and unacceptable results.
Observables are clustered into a SOM and judged by the ``user''. 
A statistical analysis of the corresponding initial-scale PDFs is 
performed and gives rise to the next iteration of PDFs.
Several criteria can be chosen by the user: from the minimization of $\chi^2$, 
to satisfying different sum rules, to selection on the behavior at low or large $x_{Bj}$, 
etc....
In this contribution we show results based on the criterion of minimization
of $\chi^2$ that allows us to gauge and test our initial results with the previously 
existing ones \cite{durham}.

\section{Method}
SOMs, at variance with 
standard NNs, are based on competitive-learning \cite{Kohonen}. 
In competitive learning one defines a number of ``filters'' that respond differently
to the initial inputs in such a way that one or few of the filters are ``winners'' producing
a high output. The ``winners'' create negative feedback so that only they and their neighbours 
get reinforced through the various cycles, or in other words, they get updated in learning.
More technically, a SOM is an algorithm that maps in a topologically ordered way the 
training data onto a neural network. 
The mapping proceeds by selecting the neuron, $N_W$, that best matches 
each data sample according to a metric, $M_D$.
Each neuron is represented in a two-dimensional grid, with coordinates:
${\bf x_i} \equiv (x_1,x_2)$.
A weighted average of each neuron, $N_i$ in the grid to the data sample
is then performed, where the weight, $w_i$ is computed from the distance
of $N_i$ to $N_W$ according to a metric, $M_G$, and a given
neighborhood radius. $M_G$ defines the topology
of the grid. 
This procedure is iterated with smaller radii until 
it saturates.

\begin{figure}[ht]
{\epsfxsize=2.1in\epsfbox{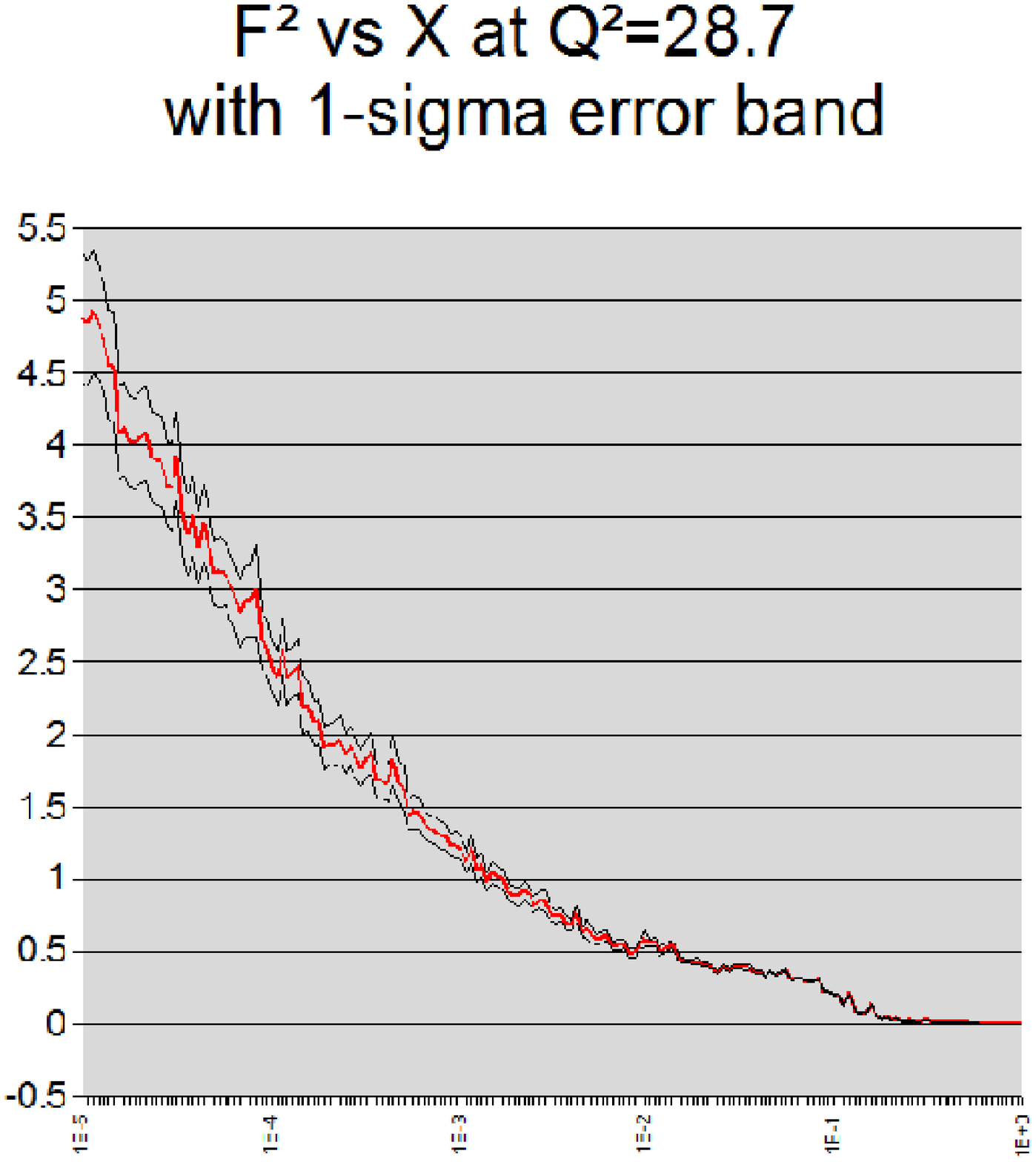}}  
{\epsfxsize=2.1in\epsfbox{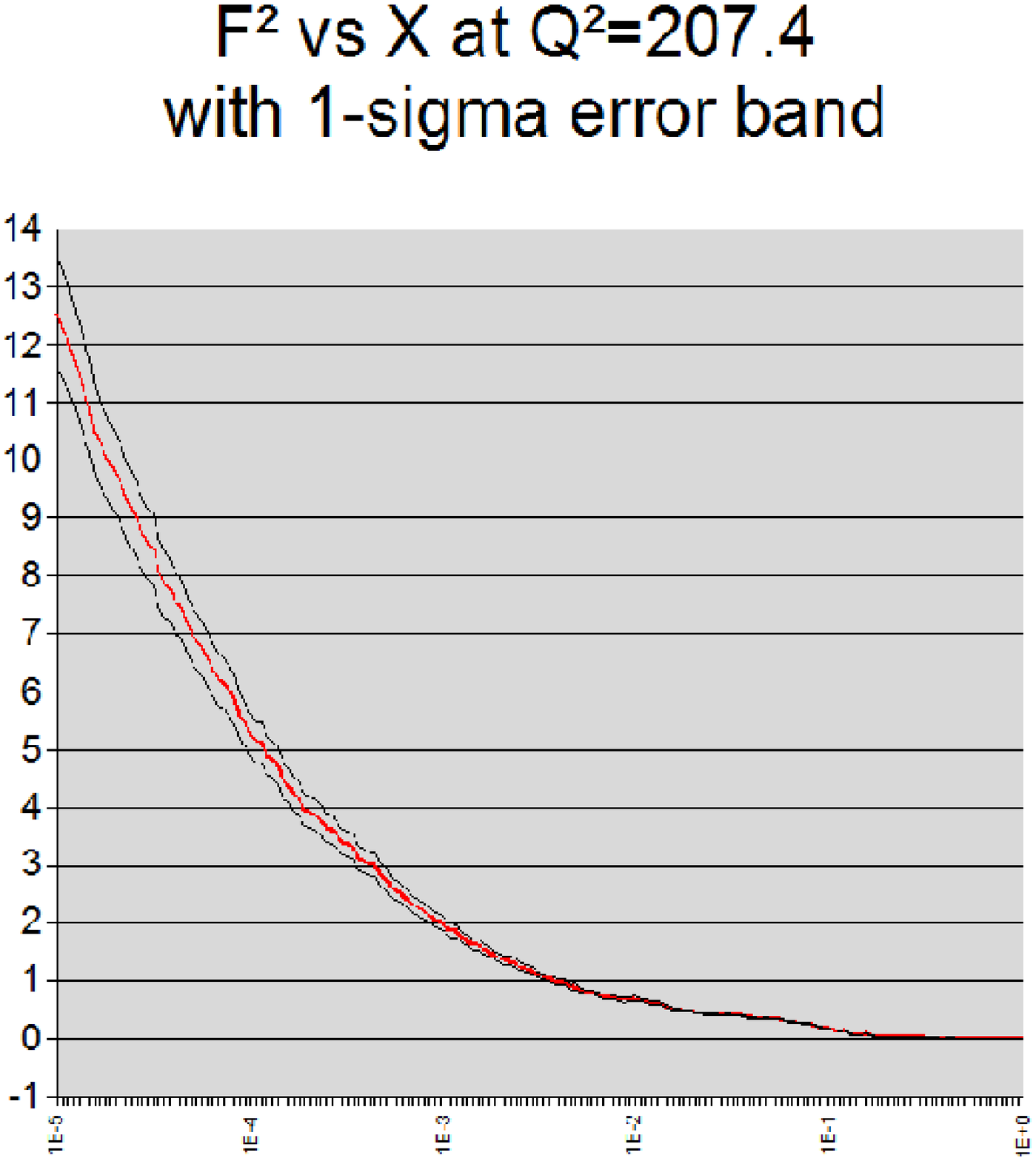}}   
\caption{Left: Structure function $F_2(x,Q^2)$ from SOMPDFs fit, plotted vs. $x$ in the range 
$10^{-5} < x < 1$, at $Q^2=28.7$ GeV$^2$; Right: $F_2(x,Q^2)$ in the same range of $x$, at 
$Q^2=207$ GeV$^2$.\label{fig1}}
\end{figure}

For our specific problem, 
the neurons correspond to the PDFs; the data are ``synthetic data''
(randomized samples of the original data). 
The metric $M_G$ that defines the topology of the map
is: 
\begin{equation}
L_1({\bf x},{\bf y}) = \sum_{j=1,2} \mid x_j - y_j \mid 
\end{equation}
An important aspect of our procedure is that PQCD evolution is considered at every step.
Our preliminary results are displayed in Fig.\ref{fig1} showing that our algorithm represents 
indeed a robust method to determine both the structure
function $F_2(x_{Bj},Q^2)$, and the gluon distribution, $G(x_{Bj},Q^2)$, evolved at
$Q^2=28.7$ GeV$^2$. 


We conclude that the proposed SOMPDFs, introudce a change of criteria with respect to NNPDFs
aimed at bringing ``theory'' back in the loop, at variance with seeking full automation of the 
fitting procedure. They are therefore placed at the intersection between traditional GF 
methods and NN approaches.
SOMPDFs have the following additional advantages over generic Genetic Algorithms
that might help in future work to identify the role of
different parameters:
{\it i)} Visualization; {\it ii)} Dimensionality reduction;
{\it iii)} Clustering (a study is on its way to determine what features of PDFs produce
given patterns of clustering).
We hope as future practical goals, to extend our investigation to addtional ``filters'' other 
than the $\chi^2$ \cite{SOMPDF1}, and to study the implementation of  
SOMPDFs in actual data analyses at the LHC
using both nucleon and nuclear data.



\begin{thebibliography}{0}
\bibitem{Pumplin:2005yf}
  J.~Pumplin,
  AIP Conf.\ Proc.\  {\bf 792}, 50 (2005)

\bibitem{durham} http://durpdg.dur.ac.uk/HEPDATA/

\bibitem{Piccione}
  L.~Del Debbio, S.~Forte, J.~I.~Latorre, A.~Piccione and J.~Rojo  [NNPDF
                  Collaboration],
  JHEP {\bf 0503}, 080 (2005), {\it and references therein}; A. Piccione, 
{\it these proceedings}.

\bibitem{cowan}
  W.~T.~Giele, S.~A.~Keller and D.~A.~Kosower,
  arXiv:hep-ph/0104052; G. Cowan, {\it these proceedings}.

\bibitem{Kohonen} T. Kohonen, {\em ``Self Organizing Maps''}, Springer-Verlag, 1997.

\bibitem{SOMPDF1} H. Honkanen, S. Liuti, Y. Loitiere, D. Brogan and P. Reynolds, 
{\it in preparation}. 

\end{thebibliography}
\end{document}